\pdfoutput=1
\documentclass[12pt]{article}
\usepackage{mathtools}
\usepackage[numbers]{natbib}
\usepackage[dvipsnames]{xcolor}
\usepackage{enumitem}
\usepackage{arxiv}
\usepackage{multirow}
\usepackage{graphicx} 
\usepackage{booktabs} 
\usepackage{amsmath}
\usepackage[utf8]{inputenc} 
\usepackage[T1]{fontenc}    
\usepackage{hyperref}       
\usepackage{url}            
\usepackage{booktabs}       
\usepackage{amsfonts}       
\usepackage{nicefrac}       
\usepackage{microtype}      
\usepackage{lipsum}
\usepackage{tabularx} 
\usepackage{algorithm}
\usepackage{algpseudocode}
\usepackage{setspace} 
\usepackage{tikz}
\usetikzlibrary{positioning}
\usetikzlibrary{shapes.arrows}
\usetikzlibrary{decorations.markings,arrows.meta}

\makeatletter
\DeclareFontFamily{OMX}{MnSymbolE}{}
\DeclareSymbolFont{MnLargeSymbols}{OMX}{MnSymbolE}{m}{n}
\SetSymbolFont{MnLargeSymbols}{bold}{OMX}{MnSymbolE}{b}{n}
\DeclareFontShape{OMX}{MnSymbolE}{m}{n}{
    <-6>  MnSymbolE5
   <6-7>  MnSymbolE6
   <7-8>  MnSymbolE7
   <8-9>  MnSymbolE8
   <9-10> MnSymbolE9
  <10-12> MnSymbolE10
  <12->   MnSymbolE12
}{}
\DeclareFontShape{OMX}{MnSymbolE}{b}{n}{
    <-6>  MnSymbolE-Bold5
   <6-7>  MnSymbolE-Bold6
   <7-8>  MnSymbolE-Bold7
   <8-9>  MnSymbolE-Bold8
   <9-10> MnSymbolE-Bold9
  <10-12> MnSymbolE-Bold10
  <12->   MnSymbolE-Bold12
}{}

\let\llangle\@undefined
\let\rrangle\@undefined
\DeclareMathDelimiter{\llangle}{\mathopen}%
                     {MnLargeSymbols}{'164}{MnLargeSymbols}{'164}
\DeclareMathDelimiter{\rrangle}{\mathclose}%
                     {MnLargeSymbols}{'171}{MnLargeSymbols}{'171}

\makeatother

\algnewcommand{\Inputs}[1]{%
  \State \textbf{Inputs:}
  \Statex \hspace*{\algorithmicindent}\parbox[t]{.8\linewidth}{\raggedright #1}
}
\algnewcommand{\Outputs}[1]{%
  \State \textbf{Outputs:}
  \Statex \hspace*{\algorithmicindent}\parbox[t]{.8\linewidth}{\raggedright #1}
}

\algnewcommand{\Functions}[1]{%
  \State \textbf{Notation:}
  \Statex \hspace*{\algorithmicindent}\parbox[t]{.8\linewidth}{\raggedright #1}
}

\newtheorem{theorem}{Principle}
\usepackage{subfigure}
\usepackage{amsmath}

\usepackage{lipsum}

\newcommand\blfootnote[1]{%
  \begingroup
  \renewcommand\thefootnote{}\footnote{#1}%
  \addtocounter{footnote}{-1}%
  \endgroup
}

\title{Composite Likelihoods with Bounded Weights in Extrapolation of Data}

\author{Margaret Gamalo, Yoonji Kim, Fan Zhang, \& Junjing Lin}

\begin{document}

\maketitle

\begin{abstract}
Among many efforts to facilitate timely access to safe and effective medicines to children, increased attention has been given to {\em extrapolation}. Loosely, it is the leveraging of conclusions or available data from adults or older age groups to draw conclusions for the target pediatric population when it can be assumed that the course of the disease and the expected response to a medicinal product would be sufficiently similar in the pediatric and the reference population. {\em Extrapolation} then can be characterized as a statistical mapping of information from the reference (adults or older age groups) to the target pediatric population. The translation, or loosely mapping of information, can be through a composite likelihood approach where the likelihood of the reference population is weighted by exponentiation and that this exponent is related to the value of the mapped information in the target population. The weight is bounded above and below recognizing the fact that similarity (of the disease and the expected response) is still valid despite variability of response between the cohorts. Maximum likelihood approaches are then used for estimation of parameters and asymptotic theory is used to derive distributions of estimates for use in inference. Hence, the estimation of effects in the target population borrows information from reference population. In addition, this manuscript also talks about how this method is related to the Bayesian statistical paradigm.  \blfootnote{Corresponding Author: Margaret Gamalo, PhD is Statistics Therapeutic Area Head, Inflammation and Immunology, Pfizer; Email:margaret.gamalo@pfizer.com. Yoonji Kim is a PhD Candidate at the Department of Statistics, The Ohio State University, Fan Zhang is Associate Director, Global Biometrics and Data Sciences, Pfizer, Junjing Lin is Associate Director, Statistical and Quantitative Sciences, Takeda. The views expressed in this paper are those of the authors and not necessarily those of the author's employer.}

\end{abstract}


\section{Introduction}
Despite the requirement and associated incentives to address the economic burdens of pediatric drug development, the typical delay in getting pediatric labeling after the initial adult approval is still an average of 9 years with many pediatric trials ending up not finishing, abandoned, or delayed \citep{mulugeta2017development,hwang2018delays}. Nearly one out of five trials ended early, primarily due to recruitment challenges, with a proportion of trials withdrawn before recruitment began \citep{pica2016discontinuation}. In a related investigation, completion of many pediatric studies required under the European Union (EU) Paediatric Regulation is generally delayed  \citep{tomasi2013paediatric}. For this reason, {\it extrapolation} is promoted as a means to reduce ``the amount of, or general need for, additional information (e.g., types of studies, design modifications, number of patients required) needed to reach conclusions’’ when it can be assumed that the course of the disease and the expected response to a medicinal product will be sufficiently similar in the pediatric (or target) and the reference (or source) population  \citep{guideline2000clinicalR1, Reflection}. This is also justified as children are considered a {\it vulnerable population}, i.e.,  children should only be enrolled in research if the scientific and/or public health goal(s) cannot be met through enrolling subjects who can consent personally \citep{roth2011ethical}.

The right number of pediatric patients also implies that extrapolation should be a default strategy in pediatric development and warrants a further use of efficient innovative designs. This includes innovative analytical strategies with appropriately designed adult clinical trials. Of note, extrapolation as a reduction of extent of development, i.e., number of trials and sample size, includes two steps: (1) determining the trial type based on the degree of information that can be translatable from the reference to the target population within the current indication and the risk associated with the development; and (2) quantitative or innovative methodologies to be implemented. These novel analytical strategies further reduce the required evidence to be obtained from the target pediatric population following the predicted degree of similarity to the source population.

The manuscript aims to offer a translation of the extrapolation concept into statistics to provide some guidance on the extent of development in a pediatric trial. In particular, it discusses  {\it composite likelihood} to borrow information from the reference population. The manuscript is organized in the following sequence. The next section discusses the concept of extrapolation as defined in the International Conference on Harmonization (ICH) E11 (R1). Section 3 discusses the composite likelihood and bounded weights. Bounded weights is a unique concept that recognizes similarity of diseases/response precedes variability in outcomes. Hence, minor changes in response, in a disease that is known to be similar in both cohorts, should not be used as penalty for down-weighting or reduction in accessible information. Section 4 provides an example of a streamlined pediatric drug development with analysis on its operating characteristics and its relationship to the concept of {\it tolerable uncertainty}. The last section gives a discussion of the key messages of the manuscript.

\begin{figure}[ht!]
     \begin{center}
      
            \label{Extrapolation_EMA_Figure}
            \includegraphics[width=5in]{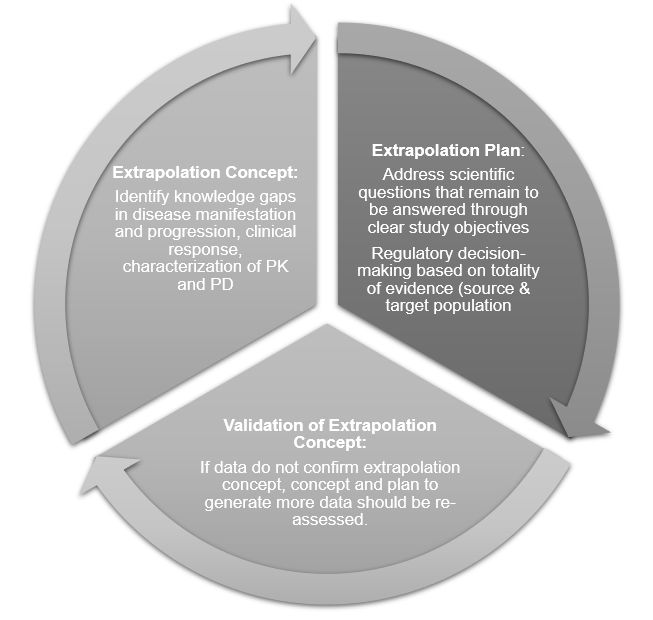}
    
    \end{center}
    \caption{Extrapolation Framework as outlined in the EMA's Reflection paper on the use of extrapolation in the development of medicines for paediatrics.
     }%
   \label{fig:PlaceboPrior}
\end{figure}

\section{Extrapolation in Pediatric Clinical Development}
In the EMA’s reflection paper, in particular, extrapolation starts with initial substantiation, i.e., assumptions forming the extrapolation concept should be evidence-based. This establishes a ``line of reasoning’’ about the relation between the disease pathogenesis or underlying cause and in tissue findings, clinical presentation or manifestation of disease, criteria for disease diagnosis and onset, disease stage or severity, co-morbidities, and treatment.'' Additional data that may help establish similarity of the disease include time course of the disease, portability of response measures, i.e., measurement can be used in target population, as well as physiologically based mathematical representations of biological, pathophysiological, and pharmacological processes in as much molecular detail as possible. 

To establish similarity in response, a systematic assessment and synthesis of available data from clinical trials is needed. Other relevant evidence from clinical practice (e.g., from other pediatric age groups, related pediatric indications, adult indications for (similar) pediatric indications, real-world evidence, or historical or placebo controls) may also provide information on the degree of similarity between adults and children in the course of the disease and the response to treatment. Reference population information to justify similarity of clinical response may come from products used to treat the same indication, other formulations of the same active ingredient, or surrogate endpoints and data from other clinical trials and observational studies. There is no universal answer to ``when is it reasonable to assume similarity of response?'' as the assessment of similarity depends on the indication, clinical consensus, and the definition of the criteria for delineating when ``similar is similar enough.''

These criteria should not be construed that the goal in the substantiation of extrapolation is to show dissimilarity between the reference population and the target pediatric population. Most research has been about measuring the right outcomes that define disease progression or emphasis on tailoring treatment options; hence, these researches have been about differences in disease than comparability. In addition, the identification of knowledge gaps can be confusing, since it focuses on disease manifestation and progression, i.e., differences between reference and target or adults and children. However, the gaps seem to be equated with knowledge of efficacy, safety, and PK. The question should be how the similarity in these parameters reduces the amount of information needed to establish, efficacy, safety, and PK. Ultimately, the mindset needs toward ensuring safe and appropriate usage in children should be emphasized rather than establishing necessary metrics of disease similarity. In the real world, clinical judgment presides over pharmacologic rationale — the dire need to provide pediatric patients the treatment they need will force a clinician to guess the appropriate administered dose once adult approval. 
 
\section{Statistical Translation of Information through Composite Likelihoods}\label{Structure}
Statistically, the conditions for applying extrapolation are stringent as similarity in disease progression implies similarity at baseline and disease prognosis. Furthermore, the criteria of similarity of treatment response imply clinical meaningful consistency of response. The challenge has often led to its limited use. However, many have argued that extrapolation should not be all or nothing and that there are degrees of extrapolation that can be accommodated \citep{guideline2022E11A}. The clinical assessment of dissimilarity on disease progression/response to intervention may not be sufficient ground for withholding treatment to a child in the light of the benefit that could be potentially achieved \citep{gamalo2022extrapolation}. Hence, some form of extrapolation or should be a default strategy in any pediatric drug development. 

In view of these arguments and to facilitate a statistical translation, suppose that we call a disease $\mathcal{D}_r$ for the reference population and when it appears in the target population as $\mathcal{D}_t$. We say $\mathcal{D}_r$ is similar to $\mathcal{D}_t$ if any information (disease progression, response to intervention) in $\mathcal{D}_r$, called $D_r$, has corresponding information in $\mathcal{D}_t$, called $D_t$. For example, a simple case would be that $\mathcal{D}_r$ similar to $\mathcal{D}_t$ if information in $\mathcal{D}_r$ is proportional to some data in $\mathcal{D}_t$. Hence, if $A$ is quantifiable evidence in $\mathcal{D}_r$ then we can determine its projected information in $\mathcal{D}_t$. Call this projection as the translatable information which is only worth $\gamma |A|$ in $\mathcal{D}_t$, for some $\gamma \in R$. In general, we are interested in a function that will map information in $\mathcal{D}_r$ to information in $\mathcal{D}_t$.

Suppose the extrapolation plan for a drug developed for $\mathcal{D}_t$ is to have sufficient information $S$ in the form of an adequate and well-controlled trial to warrant label extension in children. Since $\gamma |A|$ is already available, the objective is to conduct a trial with information size $B$ such that $\gamma|A| + |B| = |S|$. $B$ then is just the right amount of pediatric patients to fill the knowledge gap or uncertainty in the currently available data. Note that the more similar the diseases are then the less $B$ is needed to reach $S$. Furthermore, while $B$ can come from a continuum of evidence, there are only a few trial types from which information can be extracted, e.g., 
\begin{itemize}
\item PK/PD or dose-ranging study;
\item single arm descriptive efficacy and/or safety study;
\item randomized controlled efficacy and/or safety study. 
\end{itemize}From these types, the extent of the trial in terms of sample size can be increased or decreased. The reduction can be similarly based on the perceived similarity of the two diseases and can be derived based on the following principle where $\gamma$ is denoted by $w_k$ in relation to the indexed reference population $A_k$: 
\begin{theorem}
Extrapolation as a default strategy implies that information translatability $w_k$ from a reference population $A_k$ relative to the target population $A^{\ast}$ will always be $0<w_k<1$. The scalar $w_k$ is called the extrapolation coefficient of that population relative to the target population.
\end{theorem}

\subsection{Composite Likelihood}\label{Composite}
One quantitative way of incorporating the extrapolation coefficient into the analysis of treatment effects is through the use of the composite likelihood approach. Suppose that $\mathcal{D}$ has response data $\boldsymbol{Y}$ composed of $\{Y_{1},\ldots,Y_{K}\}$, where $Y_k = \{Y_{1k},\ldots,Y_{n_kk}\}$ $k=1, \ldots, K$. Let ${A_1,...,A_K}$ be population cohorts with associated likelihoods $L_k(\boldsymbol{\theta};Y)\propto f(Y\in  A_k;\boldsymbol{\theta})$. Assume that all the individuals in the cohorts have the same parameter, i.e., $\boldsymbol{\theta}_{ik}= \boldsymbol{\theta}$, $i=1, \ldots, n_k$, $k=1, \ldots, K$. Then the composite likelihood of $\boldsymbol{\theta}$ is  
\begin{equation}\label{eq:composite_lik}
CL(\boldsymbol{\theta}|\boldsymbol{Y})  \propto  \prod_{k=1}^{K}\prod_{i=1}^{n_k}L(\boldsymbol{\theta}|Y_{ik})^{w_k}.
\end{equation} 

Given that $D_t =Y_{k^*}$, where $k^*$ is the size of the target population, and $D_r = \boldsymbol{Y}\setminus D_t$, which denotes all the data except the target population, one could specify $w_{k^{\ast}}=1$ and $w_{k\neq k^{\ast}}=\alpha$, where $\alpha$ is the pre-specified {\it extrapolation coefficient} between $A_{k^{\ast}}$ and $A_{k\neq k^{\ast}}$. Then taking the logarithm of (\ref{eq:composite_lik}) yields $\sum_{i=1}^{n_{k^{\ast}}}\ell(\boldsymbol{\theta}|Y_{ik^{\ast}}) + \alpha\sum_{k\neq k^{\ast}}\sum_{i=1}^{n_k}\ell(\boldsymbol{\theta}|Y_{ik}):=c\ell(\boldsymbol{\theta}|\boldsymbol{Y})$ which has the form $ |B| + \gamma|A| = |S|$ as described previously. We will denote an asterisk for $k^{\ast}$, as ultimately the goal is to find the optimal sample size for the study in the target population. Under regularity conditions on the component log-densities, the central limit theorem for the composite likelihood score statistic leads to the result that
the composite maximum likelihood estimator, $\boldsymbol{\theta}^{CL}$, is asymptotically normally
distributed as follows. As $n=\sum_{k=1}^K n_k$ increases to infinity,
\begin{equation}\label{eq:CL-asymptotic}
\boldsymbol{\theta}^{CL}\xrightarrow{d}N_p(\boldsymbol{\theta}, \boldsymbol{G}^{-1}(\boldsymbol{\theta})),
\end{equation} where $N_p(\cdot,\ \cdot)$ is the $p$-dimensional normal distribution with mean and variance as indicated, $\boldsymbol{G}(\boldsymbol{\theta})$ is the Godambe information matrix of the whole set of observations, defined as
\begin{equation}
\boldsymbol{G}(\boldsymbol{\theta}) = \boldsymbol{H}(\boldsymbol{\theta})\boldsymbol{J}(\boldsymbol{\theta})^{-1}\boldsymbol{H}(\boldsymbol{\theta}), 
\end{equation} where $
\boldsymbol{H}(\boldsymbol{\theta}) = E_{\boldsymbol{\theta}}\{\nabla_{\boldsymbol{\theta}} u(\boldsymbol{\theta}; \boldsymbol{Y})\} =\int \{\nabla_{\boldsymbol{\theta}} u(\boldsymbol{\theta}; \boldsymbol{Y})\}f(\boldsymbol{Y}; \boldsymbol{\theta})dY$ is the Hessian matrix, $
\boldsymbol{J}(\boldsymbol{\theta}) = \mathrm{var}_{\boldsymbol{\theta}} \{u(\boldsymbol{\theta};\boldsymbol{Y})\}$
is the variability matrix, and $u(\boldsymbol{\theta};\boldsymbol{Y})=\nabla_{\boldsymbol{\theta}}c\ell(\boldsymbol{\theta}|\boldsymbol{Y})$ is the composite score function. 

Suppose that the statistical model has parameter $\boldsymbol{\theta}=(\psi, \tau)\in \boldsymbol{\Omega}$. Let $H_0: \psi=\psi_0$ and $H_1: \psi\neq\psi_0$. The composite likelihood ratio test statistic for the null hypothesis arises as
\begin{equation}
W=2\left[c\ell\left(\hat{\boldsymbol{\theta}};\boldsymbol{Y}\right) - c\ell\left(\psi_0, \hat{\tau}_{CL}(\psi_0);\boldsymbol{Y}\right)\right]
\end{equation}
which has a non-standard asymptotic distribution
\begin{equation}\label{eq:asymptotic_dist}
W\xrightarrow{d}\sum_{j=1}^q\lambda_jZ^2_j,
\end{equation} where $Z_1, \ldots, Z_q$ are independent normal variates and $\lambda_1,\ldots,\lambda_q$ are the eigenvalues of the matrix $(\boldsymbol{G}^{\psi\psi})^{-1}\boldsymbol{H}^{\psi\psi}$. Here $\boldsymbol{G}^{\psi\psi}$ is the $q\times q$ submatrix of the inverse of the Godambe information pertaining to $\psi$ and $\boldsymbol{H}^{\psi\psi}$ is the $q\times q$ submatrix of the inverse of $\boldsymbol{H}(\boldsymbol{\theta})$ pertaining to $\psi$. The Satthertwaite adjustment $W''=\nu W/(q\bar{\lambda})$, where $\bar{\lambda}$ is the average of the eigenvalues $\lambda_j$, has an  approximate $\chi^2_{\nu}$ distribution with $\nu=(\sum_{j=1}^q\lambda_j)^2/\sum_{j=1}^q\lambda_j^2$ has mean and the variance that coincides with (\ref{eq:asymptotic_dist}) which facilitates inference on $\psi$. Inference can also be carried out using the Wald-type statistic as defined as

\begin{equation}
    W_{d} = \left(\hat{\psi}_{CL}-\psi_0 \right)^\top \boldsymbol{G}^{\psi\psi}\left(\hat{\psi}_{CL}\right) \left( \hat{\psi}_{CL}-\psi_0\right),
\end{equation}
which follows $\chi^2_q$ distribution.

\subsection{Specification of Weights}\label{specification}
The leveraging of information is captured by the extrapolation coefficient $\alpha$, $0\leq \alpha\leq 1$, which is dependent on an evidence-based ``line of reasoning’’ as outlined in many extrapolation propositions (see also Hlavin in the context of increasing the testing significance levels \citep{hlavin2016evidence} and prior elicitation application as presented in \citep{ye2020bayesian}). While these proposals have been carefully thought through, refinements can also be implemented in its determination to ensure they match the intended concept. The ICH E11A ``recommends approaches to assessing factors that influence the determination of the similarity of disease and response to treatment between a reference and pediatric target population.’’ These approaches could also be used to produce a composite assessment tool. 

In particular, this assessment tool can take the form of a composite score, which can explore several factors such as epidemiology and natural history, underlying cause and tissue findings of the disease pathogenesis-genotype and phenotype, clinical presentation or manifestation of the disease characteristics, criteria for disease diagnosis, classification, onset, and disease severity, and comorbidities. In addition, the composite score can incorporate a systematic assessment and synthesis of available data from clinical trials and potentially other relevant evidence from clinical practice. Reference information, e.g., data from other clinical trials and observational studies, to justify the similarity of clinical response may also come from other indications that have the same target as the indication being studied as well as products used to treat the same indication. 

Alternatively, an overall impression of similarity of disease severity and response to therapy can be determined, for example, in terms of single domain scoring just like the prior elicitation application as presented in \citep{ye2020bayesian}). This impression can be bounded above or below recognizing uncertainties. This proposition for a similarity score can be prepared at the same time as the submission of the Pediatric Investigational Plan (PIP) and the Pediatric Study Plan (PSP) as these documents evaluate these topics qualitatively. By doing so, the overall impression or score may be pre-specified and agreed upon during the planning phase of the clinical development. Development and validation of a specific tool or decision tree for decision-making of the extrapolation coefficient will be discussed in future research.

\subsubsection{Bounding Weights Single Arm Studies} \label{bounded_single}
In the discussion above, we can represent the exponent as a function $w_k = w(Y_{k^{\ast}},Y_k)$ where $w_{k^{\ast}}=1$ and the function is pre-specified {\it extrapolation coefficient} between the target population $A_{k^{\ast}}$ and a reference population $A_{k}$ when $k\neq k^{\ast}$. Putting some flexibility on this weight, through a bound, is important because there is a possibility that even if we have set a fixed weight, the data may turn out to have more variability than initially thought. On the other hand, bounding is also necessary because there are cases where there is a clear deviation in the results and yet the chosen methodology does not quickly adjust to this information. Hence, the form of $w(Y_{k^{\ast}},Y_{k})$ can be chosen so that it adjusts when outcomes are similar and attenuates when outcomes are dissimilar but that its value is bounded in $[a,b]$. We will call this extrapolation Principle 2. This also makes the ICH E11A recommendation on understanding  ``{\it a priori} how much available information is being incorporated into the design and analysis to support the interpretation of the pediatric trial.'' The {\it a priori} quantification can be made feasible if there is a fixed amount of information that is borrowed because it is challenging if both outcomes and the degree of borrowing are not known. 

\begin{theorem}
Suppose the measurement of disease dissimilarity between the target population $A_{k^{\ast}}$ and a reference population $A_{k},\ k\neq k^{\ast},$ is $\tau_k$, which is derived from the outcomes $Y_{k^{\ast}}$ and $Y_k$ with value 0 representing minimal or none heterogeneity and larger $|\tau_k|$  representing larger heterogeneity. To account for heterogeneity within context of similarity, the weights can be replaced by a function $w:\tau_k\rightarrow w(\tau_k)$, where $a\leq w(\cdot)\leq b\subset (0,1)$ and the $w$ function is often defined as a bell shaped function with maximum at 0. 
\end{theorem}

One candidate group of functions that can be used for $w(\cdot)$ are variations of the so-called bounded influence functions and their weight functions. For example, the Huber weight function $w_H(\tau)=I(|\tau|\leq c)+\frac{c}{|\tau|}I(|\tau|\geq c)$ and the bisquare weight function $w_b(\tau) = (1-(\tau/c)^2)^2I(|\tau|\leq c)$, where $c$ is a tuning constant,  have the desired behavior. This can be modified in a few different approaches, in the context of leveraging information, so that the maximum weight $b$ happens when the similarity measurement between the reference and target populations is high enough and the weight decreases as the extent of estimated dissimilarity between two populations increases and is bounded by a pre-specified minimum value $a$.

\begin{figure}[ht!]
     \begin{center}
        \subfigure[Weight function $w_1$]{%
            \includegraphics[width=2.5in]{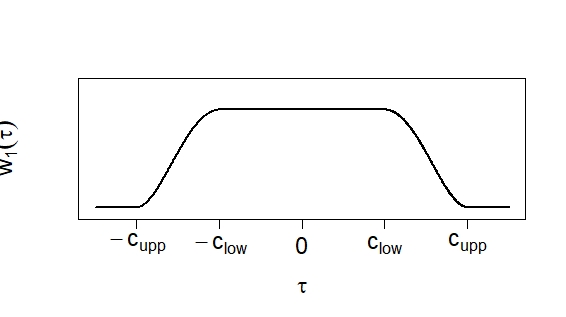}
        }%
        \subfigure[Weight function $w_2$]{%
            \includegraphics[width=2.5in]{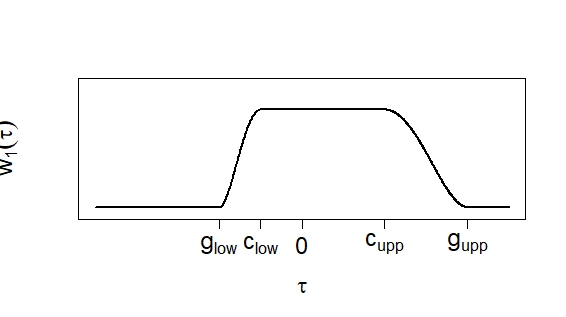}
        }\\
    \end{center}
    \caption{Visualization of weight functions which are modified versions of biweight and Huber weight functions.}%
   \label{fig:weight_functions}
\end{figure}

The Huber function has a pragmatic clinical interpretation. When this is applied as a weight function for borrowing information, the rate being borrowed will be constant where the two population response rates are similar (or within an interval) and decreases outside that interval. Motivated by desirable properties of these two existing weight functions, a weight function $w_{1}$ is defined as below.
\begin{equation}
   w_1 (\tau) =     
        \begin{cases}
          b & |\tau| < c_{low},\\
          a+(b-a)\left(1-\left(\frac{\tau-c_{low}}{c_{upp}-c_{low}}\right)^2 \right)^2 & c_{low} \leq |\tau| \leq c_{upp},\\
          a & |\tau| > c_{upp},
        \end{cases} 
\end{equation}
where the upper and lower bounds of the domain $c_{low}$ and $c_{upp}$ is pre-specified based on the application and making the weight function more clinically meaningful. The proposed function is visualized in Figure \ref{fig:weight_functions}. While we adopt the constant upper bound of the Huber weight function, we apply the overall shape of the bisquare weight function since it decreases smoothly beyond the interval $[-c_{low},c_{low}]$ and hits the minimum value $a$ beyond the interval $[-c_{upp},c_{upp}]$. The Huber weight function, on the other hand, decreases but never hits the minimum value. A set minimum weight $a$ is used when the dissimilarity statistic between the reference and target population is at least a certain value called the tuning constant $c_{upp}$. As the dissimilarity increases, the weight smoothly increases from a threshold $c_{upp}$ and the weight has the maximum value $b$ when the response in the target population is close enough to the reference population, i.e. when the dissimilarity measurement is less than a threshold $c_{low}$.

The dissimilarity can be measured through a difference between statistics of two groups of observations that is clinical of interest. A statistic that can be generally applied is a sample average $\tau_m(Y_{k^\ast},Y_k):=\frac{1}{n_{k^\ast}}\sum_{i=1}^{n_{k^\ast}}Y_{k^\ast,i}-\frac{1}{n_{k}}\sum_{i=1}^{n_{k}}Y_{k,i}$ (or a sample average of a sufficient statistic).
When applying a symmetric weight function such as $w_1,$ the weight is the same when the absolute difference between the sample averages of pediatric and adult trial outcomes is the same no matter which one is greater than the other. The weight function $w_1$ may be extended to consider asymmetric weight dpending on which group has a greater statistic than the other.
\begin{equation}
   w_2 (\tau) =     
        \begin{cases}
          a & \tau < g_{low},\\
          a+(b-a)\left(1-\left(\frac{\tau-c_{low}}{c_{upp}-c_{low}}\right)^2 \right)^2 & g_{low} < \tau \leq c_{low},\\
          b & c_{low} \leq \tau \leq c_{upp},\\
          a+(b-a)\left(1-\left(\frac{\tau-c_{upp}}{g_{upp}-c_{upp}}\right)^2 \right)^2 & c_{upp} < \tau\leq g_{upp},\\
          a & \tau > g_{upp},
        \end{cases} 
\end{equation}
where the intervals are pre-specified according to the application and dissimilarity measurement $\tau$.

Another weight specification approach is to adopt the p-value-based weight specification as proposed by Liu for Bayesian power priors, which is also known as the test-then-pool approach \cite{liu2018PP}. As the goal of weight specification is to increase the weight when outcomes of two groups are similar and attenuate the borrowing when outcomes are dissimilar, the p-value may be used to measure the congruence between two groups. The quantity $\tau$ is then defined as the unpaired two-sample $t$-test statistic or the Welch's t-test statistic for unequal variance and denotes the corresponding p-value by $p_v(\tau)$, where the null hypothesis is that the mean (or the mean of sufficient statistic) of two populations are the same. To smooth the weight function concerning the congruence, the weight function is defined as below.
\begin{equation}
    w_3(\tau) = a+(b-a)\ \exp\left(\frac{c}{p_v(\tau)}\log \left(1-p_v(\tau)\right)\right),
\end{equation}
where $c$ is a pre-specified shape parameter and the weight function is bounded by $[a,b]$. The full implementation steps are as follows: (1) the weights are first estimated through the weight functions described in this section; (2) given the weight estimates, the maximum likelihood estimate (MLE) of $\boldsymbol{\theta}$ is derived. This is also applicable to the multi-arm case. 

\subsubsection{Bounding Weights Multi-Arm Studies}  \label{bounded_multiple}
Denote $Y_{k} = (Y_{kt},Y_{kp})$, where $Y_{kt}$ and $Y_{kp}$ are the data for treatment group and control group from $k$th cohort respectively. There are several approaches to the extent the methods and weight function $w$ specified in Section \ref{specification}. Of note, while the discussion herein is for a two arm trial, the methods can be conveniently extended to multi-arm trials. 

Option one is to specify a separate weight function and derive the weight separately for $Y_{kt}$ and $Y_{kp}$. This approach is easy to interpret. One special case of interest is to set the maximum amount of borrowing to be  $(b_t, b_c)$ for the treatment and control group, with $( b_c > b_t)$ or $b_t = 0$ . It reflects that there is a stronger belief in the similarity of response in the control group and the preference for the response in the treatment group to be driven by the observed data in the target population.

Option two is to define the heterogeneity based on the difference in response for active and control groups. When using a simple average, $\tau_m(Y_{k^\ast},Y_k):= |(\overline{Y}_{k^\ast t} - \overline{Y}_{k^\ast p}) - (\overline{Y}_{kt} - \overline{Y}_{kp}) | $. A p-value based weight specification will be based on testing the equality of treatment difference in the two cohorts $H_0: (\theta_{k^\ast t} -  \theta_{k^\ast p}) - (\theta_{kt} -  \theta_{kp}) =0$. This method is preferred when there is a strong belief in the constant treatment difference in different cohorts.

Option three is to define the overall heterogeneity. When using a simple average, $\tau_m(Y_{k^\ast},Y_k): = |\overline{Y}_{k^\ast t} - \overline{Y}_{kt} | + |\overline{Y}_{k^\ast p} - \overline{Y}_{kp} |$. P-value based weight specification will be based on testing $H_0: \theta_{k^\ast t} = \theta_{kt} $ and $ \theta_{k^\ast p} = \theta_{kp} $.

\subsection{Bounding and Intrinsic Validation and Consistency}
Once the trial has been conducted, the last step in extrapolation is to check whether the observation derived from $B$ is consistent with what was seen in $A$ in the reference population \citep{Reflection}. If they are, then extrapolation has been validated in the sense that the response to treatment is empirically similar. In practice, if the trial shows benefit in the target pediatric population, validating the similarity of treatment responses is just an ancillary step.

In controlled clinical trials, this step has implications on (1) whether using the translatable information is indeed warranted; and also because (2) the degree of inconsistent information also has impact on the extent of nominal data needed, i.e., the reference data can heavily influence the conclusion. Note that inconsistency is {\it post facto} or after the fact - did we get it very wrong? For example, a trial in the target population  has a high placebo response compared to the reference population. Then using the translatable information in placebo from the reference population will underestimate the placebo response and make it possible for the drug to look efficacious even if there is no difference seen in the clinical trial. While it is possible that this observation could have been an artifact of smaller sample size, the posterior distribution of placebo, which gives a unified probability statement of diverging effects, would still reveal that the probability of an effect is small and a casual observer would still think that the drug is at worst not efficacious in children. The proposed methodology as described in Section \ref{bounded_multiple} embeds a check consistency and subsequently validation. 

\subsection{Analog Bayesian Specification}\label{CPP}

Gamalo-Siebers et al. \citep{gamalo2016statistical} described several ways of incorporating information from the trial in the reference population into a new trial in the target population within the extrapolation framework. In particular, some of the modeling methods, e.g., conditional power prior \citep{ibrahim2000power} and commensurate power prior \citep{hobbs2011hierarchical} described in their work can be seen as analog formulation of the composite likelihood $CL(\boldsymbol{\theta})$. As a simple example previously described, suppose that the target data $\mathcal{D}_t$ is composed of $A_{k^{\ast}}$ and has outcomes $\boldsymbol{Y}_{k^{\ast}}$ and the reference data $\mathcal{D}_r$ is composed of $A_{k}$ where $k\neq k^{\ast}$. A simple Bayesian hierarchical model for the data is for each $Y_{ik^{\ast}}$, $i=1, \ldots, n_{k^{\ast}}$, to be distributed $F$ (or density $f$) with patient level parameter $Y_{ik^{\ast}}$, i.e.,$Y_{ik^{\ast}}|\boldsymbol{\theta}_{ik^{\ast}} \sim F(\cdot|\boldsymbol{\theta}_{ik^{\ast}}, \sigma), \quad \vartheta_{ik^{\ast}}=g(\boldsymbol{\theta}_{ik^{\ast}})|\boldsymbol{\eta} \sim H(\boldsymbol{\eta}), \quad \boldsymbol{\eta}\sim Q$. $\vartheta_i$ for each patient can be described by a distribution $H(\boldsymbol{\eta})$ through a link function $g$ and the common population parameter of interest $\eta$ which has an initial prior described by $Q$. Suppose, on the other hand, the reference population has a similar model so that $\vartheta_{ik}=g(\boldsymbol{\theta}_{ik})|\boldsymbol{\eta} \sim H(\boldsymbol{\eta}) $, $\boldsymbol{\eta} \sim Q$. This model contains information on the efficacy of the drug in the reference population. The goal is to quantify the relevant information in this data with respect to how similar $D_t$ is to $D_r$ to reduce the required data in the target pediatric population. Hence, we are interested in the posterior distribution of  $\boldsymbol{\eta}$ that is sufficient to make statements about the efficacy of the drug in the target pediatric population. That posterior is simply given by the likelihood of $\boldsymbol{Y}_{k^{\ast}}$ given $\boldsymbol{\eta}$ multiplied by the prior of $\boldsymbol{\eta}$, i.e., 
\begin{equation}\label{eq:posterior0}
\pi(\boldsymbol{\eta}|\boldsymbol{Y}_{k^{\ast}})  \propto  \prod_{i=1}^{n_{k^{\ast}}} L(\boldsymbol{\vartheta}_i|Y_{ik^{\ast}}, \boldsymbol{\eta}) \pi(\boldsymbol{\eta})
\end{equation} where $\pi(\boldsymbol{\eta})$ is derived from information from the reference population given by 
\begin{equation}\label{eq:power_prior}
\pi(\boldsymbol{\eta}|\boldsymbol{Y}_{k}, \alpha) \propto \prod_{k\neq k^{\ast}} \prod_{i=1}^{n_{k^{\ast}}} L(\boldsymbol{\vartheta}_i|Y_{ik}, \boldsymbol{\eta})^{\alpha} \pi(\boldsymbol{\eta})
\end{equation} The parameter $\alpha$ governs the impact of the reference data on the current target data analysis, ranging from no influence when $\alpha=0$ to parity when $\alpha=1$. We note also that the (\ref{eq:power_prior}) will be proper as long $C\equiv\int \prod_{k\neq k^{\ast}} \prod_{i=1}^{n_{k^{\ast}}} L(\boldsymbol{\vartheta}_i|Y_{ik}, \boldsymbol{\eta})^{\alpha} \pi(\boldsymbol{\eta})d\boldsymbol{\eta}<\infty$. Combining (\ref{eq:posterior0}) and (\ref{eq:power_prior}) yields
\begin{equation}\label{eq:posterior1}
\pi(\boldsymbol{\eta}|\boldsymbol{Y}_{k}, \alpha) \propto \prod_{k=1}^K \prod_{i=1}^{n_{k^{\ast}}} L(\boldsymbol{\vartheta}_i|Y_{ik}, \boldsymbol{\eta})^{w_k} \pi(\boldsymbol{\eta})
\end{equation} where $w_k=\alpha$ for $k\neq k^{\ast}$ and $w_k=1$, otherwise. The posterior distribution is of the same form when using bounded weights since these weights are still incorporated conditionally. If they are not specified conditionally, then the posterior distribution is similar to a modified power prior where the distribution of the weight is constrained to a very narrow band, i.e., 
\begin{equation}\label{eq:modified_posterior}
\pi(\boldsymbol{\eta},\alpha|\boldsymbol{Y}_{k^{\ast}}) \propto \prod_{k=1}^K \prod_{i=1}^{n_{k^{\ast}}} L(\boldsymbol{\vartheta}_i|Y_{ik}, \boldsymbol{\eta})^{\alpha} \pi(\boldsymbol{\eta}, \alpha)
\end{equation} where if $\pi(\boldsymbol{\eta}, \alpha) = \pi(\boldsymbol{\eta})\pi(\alpha)$ then the support of $\pi(\alpha)$ is over a narrow range $(m_0, m_0+m_1)\subset (0,1)$. Hence, the approach is equivalent to having a weight equivalent to the mean of a truncated distribution of $\alpha$. 

The composite likelihood and the conditional power prior have a similar interpretation of treatment effects when the prior used for the latter is vague. The only difference is in the estimation where the composite likelihood uses estimates that maximizes the likelihood in contradistinction to using the mean of the distribution. The use of bounded weights does not change the interpretation as these weights are still specified conditionally.
When the hyperpriors are informative, the interpretation of the treatment effect may have a slightly different interpretation as it is already influenced by a prior belief. If this prior is non-informative, i.e., not based on data then the effect is only influenced by a mixture of information coming from data in the reference population calibrated by the extrapolation coefficient, data from the target population, and data from the elicitation of the prior. Hence care is needed in the interpretation.  

\section{Investigations on Operational Characteristics of the Method}
\subsection{General Exponential Family} 
Consider a one-dimensional exponential family of probability distributions given a measure $\theta$, which generally describes most endpoints used in clinical trials, with the following form.
\begin{equation}
  f(y\mid\theta) = h(y) \exp{(\theta\ T(y) - A(\theta))}.
\end{equation}
The corresponding composite log-likelihood is as follows:
\begin{equation}
c\ell(\theta|\boldsymbol{Y})  \propto  \sum_{k=1}^{K}w_k \left(\theta\sum_{i=1}^{n_k}T\left(Y_{ik}\right)-n_k A(\theta)\right).
\end{equation}
Given the weights specified as described in Section \ref{bounded_single}, the maximum likelihood estimate (MLE) can be solved in closed form by extending the property of exponential family, $\frac{\partial}{\partial \theta}A(\theta) = \mu(\theta),$ where $E(T(Y))$ is denoted by $\mu(\theta)$ for $Y\sim f$ and is a function of $\theta.$
\begin{equation}
    \widehat{\mu(\theta)} = \frac{\sum_{k=1}^{K}w_k\sum_{i=1}^{n_k}T(Y_{ik})}{\sum_{k=1}^{K} w_k n_k},
\end{equation}
The MLE of $\theta$ is then derived by applying the inverse mapping of $\mu$,
\begin{equation}
    \widehat{\theta} = \mu^{-1}\left(\frac{\sum_{k=1}^{K}w_k\sum_{i=1}^{n_k}T(Y_{ik})}{\sum_{k=1}^{K} w_k n_k}\right).
\end{equation}

\subsection{Weight and Parameter Estimation: Single Arm Binomial Trial} \label{sec:sim_study}
\begin{figure}[!t]
  \centering
  \begin{tabularx}{15cm} { 
  |c|>{\centering\arraybackslash}X |
   >{\centering\arraybackslash}X| }
    \hline
    &$n_k=800$ & $n_k=100$
    \\\hline
    \rotatebox[origin=c]{90}{Weight}&\begin{minipage}{7cm}
    \centering\includegraphics[width=7cm]{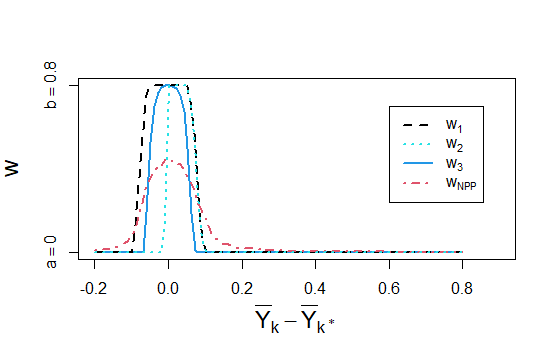}
    \end{minipage}
    &\begin{minipage}{7cm}
    \centering\includegraphics[width=7cm]{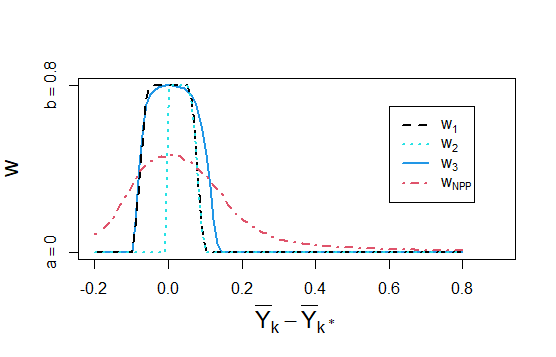}
    \end{minipage}
    \\\hline
    \rotatebox[origin=c]{90}{Estimate of $p$}&\begin{minipage}{7cm}
    \centering\includegraphics[width=7cm]{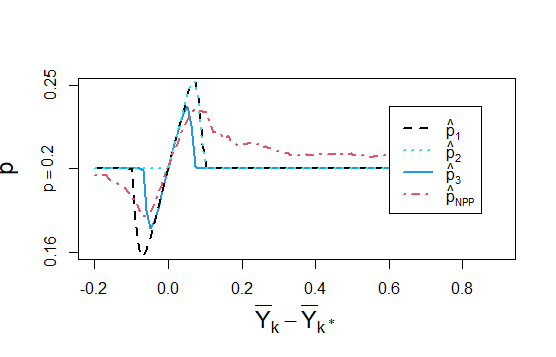}
    \end{minipage}
    &\begin{minipage}{7cm}
    \centering\includegraphics[width=7cm]{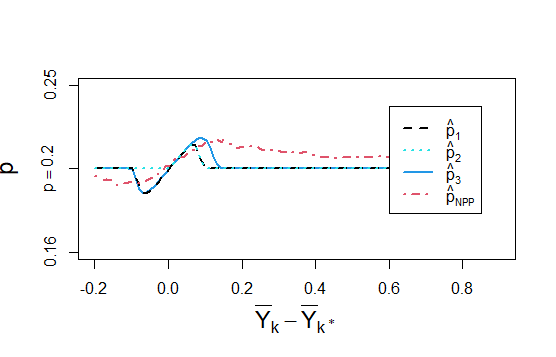}
    \end{minipage}
    \\\hline
    \rotatebox[origin=c]{90}{P-value (LRT)}&\begin{minipage}{7cm}
    \centering\includegraphics[width=7cm]{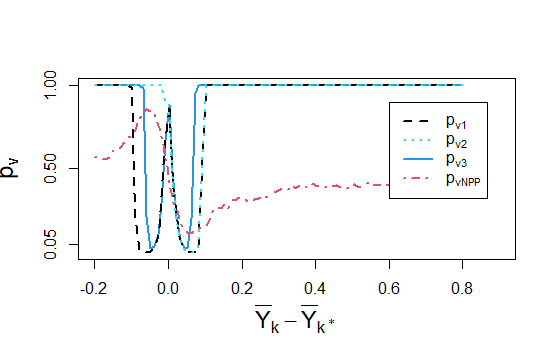}
    \end{minipage}
    &\begin{minipage}{7cm}
    \centering\includegraphics[width=7cm]{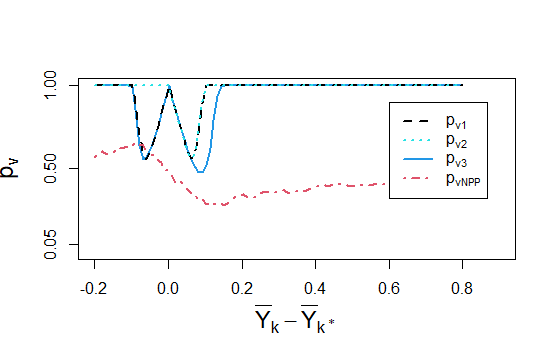}
    \end{minipage}
    \\\hline
  \end{tabularx}
  \caption{Illustration of weight, estimates of $p$, and p-values as the sample average of reference observations change fixing the number of reference observations as $800$ (column 1) and $100$ (column 2).}
    \label{fig:ex1_fixn0}
\end{figure}

\begin{figure}[!t]
  \centering
  \begin{tabularx}{15cm} { 
  |c|>{\centering\arraybackslash}X |
   >{\centering\arraybackslash}X| }
    \hline
    &$\overline{Y}_k=0.14$ & $\overline{Y}_k=0.26$
    \\\hline
    \rotatebox[origin=c]{90}{Weight}&\begin{minipage}{7cm}
    \centering\includegraphics[width=7cm]{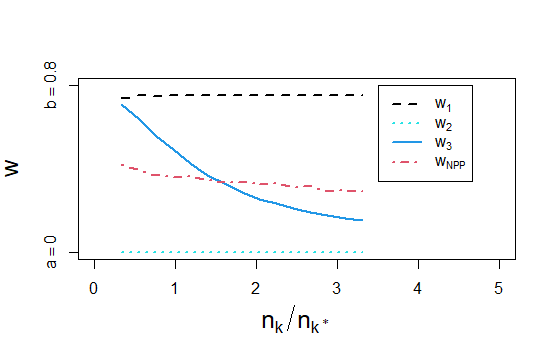}
    \end{minipage}
    &\begin{minipage}{7cm}
    \centering\includegraphics[width=7cm]{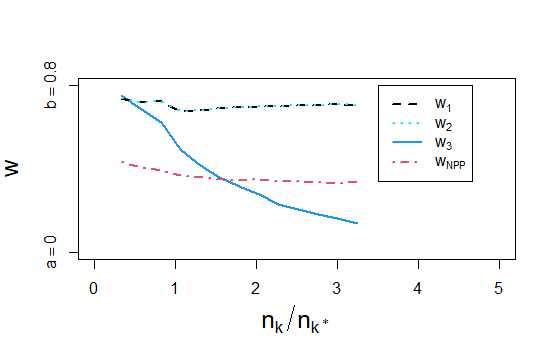}
    \end{minipage}
    \\\hline
    \rotatebox[origin=c]{90}{Estimate of $p$}&\begin{minipage}{7cm}
    \centering\includegraphics[width=7cm]{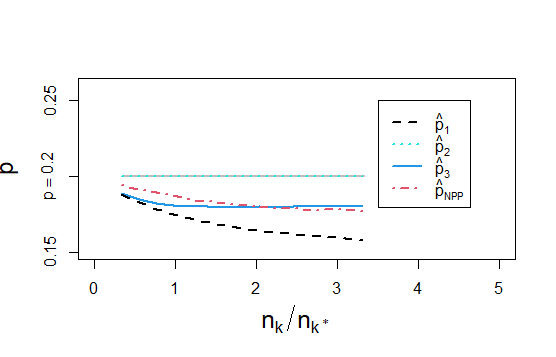}
    \end{minipage}
    &\begin{minipage}{7cm}
    \centering\includegraphics[width=7cm]{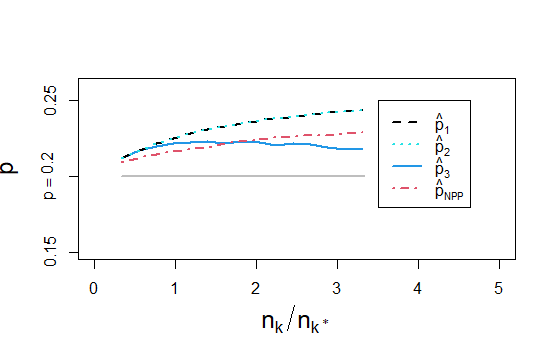}
    \end{minipage}
    \\\hline
    \rotatebox[origin=c]{90}{P-value (LRT)}&\begin{minipage}{7cm}
    \centering\includegraphics[width=7cm]{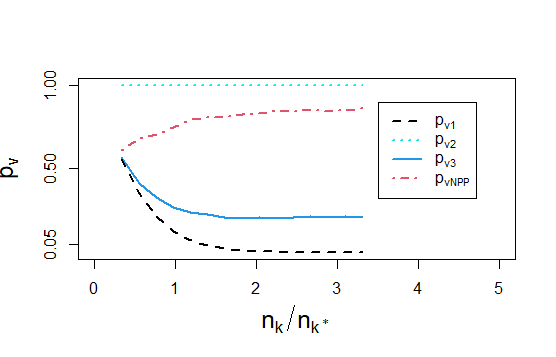}
    \end{minipage}
    &\begin{minipage}{7cm}
    \centering\includegraphics[width=7cm]{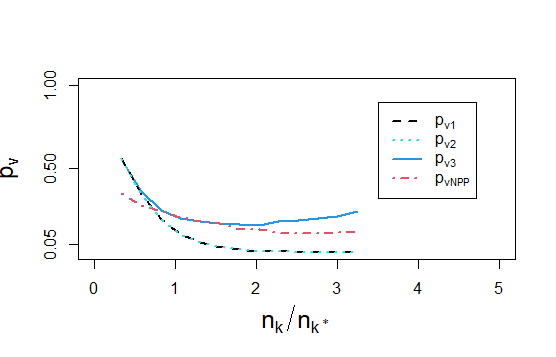}
    \end{minipage}
    \\\hline
  \end{tabularx}
  \caption{Visualization of weight, estimates of $p$, and p-values as the number of reference observations change fixing the sample average of reference observations as $\overline{Y}_k=0.14$ (column 1) and $\overline{Y}_k=0.26$ (column 2).}
    \label{fig:ex1_fixD0}
\end{figure}

In this section, we explore a binomial response model, which is a one-dimensional exponential family where $T(y) = y,\ \theta = p/(1-p),\ A(\theta) = \log(1+e^\theta)$. We assume that there is one target population with sample size $k^*$ and one reference population with sample size $k$. In Figure \ref{fig:ex1_fixn0} and \ref{fig:ex1_fixD0}, over 50 simulated datasets were generated. In these simulated datasets, we fix the target observations and we generate different reference observations from a fine grid of points on the domain $\bar{Y}-\bar{Y}_{k^{\ast}}$ to show how the estimation results vary with the different degree of dissimilarity between the target and reference observations on the response variable. For a given number of observations, we generate a dataset so that the sample average is fixed, rather than fixing the population average and randomly drawing observations from the corresponding population distribution. For example, if we set the sample average to be 0.2 and the number of observations to be 300, we would have 60 number of ones and 240 number of zeros in one dataset. This allows us to demonstrate the behavior of the proposed framework without any randomness. Given the target and reference observations, suppose the agreement with regulators was that the weights are bounded at $[0, 0.8]$, i.e., the maximum borrowing is 0.8 (see Section \ref{specification}), through functions $w_1,\ w_2$, and $w_3$ presented in Section \ref{bounded_single}. For $w_1$, let $c_{low}=0.05$ and $c_{upp}=0.1$. For $w_2,$ we set $g_{low}=-0.01,\ c_{low}=0,\ c_{upp}=0.05,$ and $g_{upp}=0.1.$ For $w_3,$ let $k=0.01.$ We then estimate $p$ given the weights, and derive the 95\% CI using $\boldsymbol{H}=((n_{k^*}+w_kn_k))/(p(1-p))$ and $\boldsymbol{J}=((n_{k^*}+w_k^2n_k))/(p(1-p))$.


We also perform Bayesian inference through a normalized power prior (NPP) where the prior models are $p \sim U[0,1]$ and $w_k \sim U[0,0.8]$ and report point estimates (marginal posterior mean) \cite{ye2022npp}. For Bayesian inference, we first obtain $1,000$ draws from the posterior distribution of $p$ and derive the 95\% posterior credible intervals.

Figure \ref{fig:ex1_fixn0} illustrates the change of weight and estimates of $p$ and p-values as the sample average of reference observations change while fixing the number of reference observations as $800$ (column 1) and $100$ (column 2).
We denote two sample averages by $\overline{Y}_k:=n_k^{-1}\sum_{i=1}^{n_k}Y_{in_k}$ and $\overline{Y}_{k^*}:=n_{k^*}^{-1}\sum_{i=1}^{n_{k^*}}Y_{in_{k^*}}$. While the weight functions with regard to the difference between two sample averages $\overline{Y}_k-\overline{Y}_{k^*}$ are invariant to the number of observations for $w_1$ and $w_2,$ the p-value based weight function $w_3$ and the marginal posterior weight $w_{NPP}$ from NPP have a sharper peak at $\overline{Y}_k-\overline{Y}_{k^*}=0$ when then number of reference observations is greater which is expected. The shape of weight functions may be adjusted by changing certain quantities in $w_1,$ $w_2,$ and $w_3$ functions allowing for a more gradual ascent and descent. It is also apparent here that the NPP does not borrow much even if there is no difference in the mean response. 

The estimates of $p$ are close to the ground truth value $p=0.2$ when $\overline{Y}_k-\overline{Y}_{k^*}=0$ or when the sample average difference is large ($> 0.1$) as this results in no borrowing. It is skew-symmetric about $(\overline{Y}_k-\overline{Y}_{k^*}=0,\ p=0.2)$ because of the influence of borrowing depending on which side $\overline{Y}_k$ is in relation $\overline{Y}_{k^*}$. They increase as the reference sample average increases and estimates of $p$ tend to increase sharply up to a peak, decrease down to $0.2$, and then are constant. The peak and trough are resulting from the dynamic borrowing of reference information with different sample averages. Of note, the smoothness and rate of increase also depend on the shape of the weight functions. While the extent of borrowing decreases accounting for the difference in two groups of data, it is relatively high when the difference in sample average is in a certain interval, resulting in the peak and trough, and is zero as we set the lower bound to be $a=0$, resulting in a constant estimate. As a small difference in sample averages is interpreted as two populations being similar, the borrowing strength is strong in such a case. Comparing different numbers of reference observations, estimates of $p$ are closer to the ground truth value (smaller peak and trough) when the number of reference observations is smaller as the proportion of reference observation affects the $p$ estimate.

When there is a large number of reference observations $n_k=800$, the p-values are large ($> 0.1$) whenever the sample average of two groups of observations are very similar ($|\bar{Y}_k-\bar{Y}_{k^*}|<0.02$). Of note, the null hypothesis in this investigation is set as the sample average of target observations. As the difference in sample average increases, the p-values decrease exponentially to a trough ($< 0.05$). This is because the sample average of reference observations shifts from the null hypothesis while the difference in sample average is small enough resulting in a moderate amount of borrowing from the reference information. When the difference between two groups of observations gets greater, this results in a small amount of or no borrowing and an increase in the p-values. For a smaller number of reference observations, e.g., $n_k=100,$ the overall pattern is similar with the lowest p-value being higher because there is less number of observations that can be borrowed.

We also explore simulated data by fixing the number of target observations $n_{k^*}=300$ and the sample average of target observations, $\overline{Y}_{k^*}=0.2$ and reference observations $\overline{Y}_k=0.14$ or $\overline{Y}_k=0.26$ and varying the number of reference observations $n_k$. 
Figure \ref{fig:ex1_fixD0} illustrates the weight, estimates of $p$, and p-values as in Figure \ref{fig:ex1_fixn0}.
When the sample average of reference observations is smaller or greater than that of the target observations by 0.06, the weights specified by $w_1$ and $w_2$ functions tend to be constant across the number of reference observations and those specified by $w_3$ and $w_{NPP}$ tend to decrease as the number of reference observations increase.
The weights specified by the $w_2$ function are smaller when the sample average of the reference group is smaller than that of the target group as the $w_2$ function is asymmetric about $\overline{Y}_k-\overline{Y}_{k^*}=0$.
The estimates of $p$ deviate far from $p=0.2$ as the number of reference observations increases and the degree of divergence is proportional to the weights shown in the first row of Figure \ref{fig:ex1_fixD0}.
For a hypothesis test, while the ground truth probability of the target population is the same as the null hypothesis $p=0.2,$ the p-values are decreasing and below the significance level when the weight and the number of reference observations are high.
When $\overline{Y}_k=0.14$, the p-values when using $w_2$ function are higher since $w_2$ is asymmetric and those for Bayesian inference are high and increasing with $n_k$ as it is performing one-sided testing.

\section{Application }\label{Augmentation}
Suppose $y_{ik}$ has sampling distribution $y_{ik}|\theta_{ik}, \boldsymbol{x}_{ik} \sim F(\theta_{ik};\boldsymbol{x}_{ik}, n_{ik})$, and the instances with which the $\theta_{ik}$, $i=1, \ldots, n_k$, occur are indistinguishable conditional on $\boldsymbol{x}_{ik}$ (i.e., the $\theta_{ik}$ are {\em exchangeable}), then, for instance, the treatment response model can take the form of 
\begin{equation}\label{eq:workingmodel}
\vartheta_{ik}=b_0 + b_1 I_{z_{ik}}(1) + \boldsymbol{x}_{ik}^{\top}\boldsymbol{\gamma},
\end{equation} where $z_{ik}$ denotes a binary covariate indicating the usage of treatment. In the example, the covariates that are thought to be effect modifiers and are thus included in the model are region, previous non-biologic therapy, and weight group. Define  $\boldsymbol{\eta}\equiv (b_0, b_1,\boldsymbol{\gamma}^{\top})$ where $b_0$ is the response of control, $b_1$ is the effect of the investigational treatment relative to control, and $\boldsymbol{\gamma}$ is a vector of incremental change per unit change in a set of covariates, all in the logit scale. If $g(\cdot)$ is the logit function and $\vartheta_{ik}=X_{ik}^{\top}\boldsymbol{\eta}$ then the composite likelihood has the form
\begin{equation}\label{eq:likelihood_ex}
q(\boldsymbol{\theta}|\boldsymbol{A}_1, \ldots, \boldsymbol{A}_K) = \prod_{k=1}^{K}w(\hat{p}_{k^{\ast}}, \hat{p}_{k}) \prod_{i=1}^{n_k}  \mathrm{exp}\left(y_{ik}(X_{ik}^{\top}\boldsymbol{\eta})-\mathrm{log}(1+ \mathrm{exp}(X_{ik}^{\top}\boldsymbol{\eta}))\right).
\end{equation} Suppose $w(\hat{p}_{k^{\ast}}, \hat{p}_{k})$ is a fixed or approximated from a dissimilarity between population $k^*$ and $k$ prior to estimation of the parameter $\boldsymbol{\eta}$.
Then, the matrices $\boldsymbol{H}$ and $\boldsymbol{J}$ are as below. Denoting $\exp({X_{ik}^{\top}\boldsymbol{\eta}})/(1 + \exp({X_{ik}^{\top}\boldsymbol{\eta}}))^2$ by $S(X_{ik}^{\top}\boldsymbol{\eta})$,
\begin{equation}
    \boldsymbol{H}(\boldsymbol{\eta}) = \sum_{k=1}^K\sum_{i=1}^{n_k}w(\hat{p}_{k^{\ast}}, \hat{p}_{k})X_{ik}S\left(X_{ik}^{\top}\boldsymbol{\eta}\right)X_{ik}^{\top},
\end{equation}
\begin{equation}
    \boldsymbol{J}(\boldsymbol{\eta}) = \sum_{k=1}^K\sum_{i=1}^{n_k}w(\hat{p}_{k^{\ast}}, \hat{p}_{k})^2X_{ik}S\left(X_{ik}^{\top}\boldsymbol{\eta}\right)X_{ik}^{\top},
\end{equation}
As presented in Section \ref{Composite}, the asymptotic covariance matrix of MLE of $\boldsymbol{\eta}$ is then $\boldsymbol{H}(\boldsymbol{\eta})^{-1}\boldsymbol{J}(\boldsymbol{\eta})\boldsymbol{H}(\boldsymbol{\eta})^{-1}.$

Note that $p_{k^{\ast}}$ and $p_{k}$ may be defined as the probability of achieving response under the investigational treatment, i.e., $E[Y_{1k^\ast}]$ and $E[Y_{1k}]$.
For the numerical study, we estimate two quantities using the sample average probability of treatment being effective for each subgroup.
$w(\hat{p}_{k^{\ast}}, \hat{p}_{k})$ is then derived using three weight functions defined in Section \ref{bounded_single} prior to estimating $\boldsymbol{\eta}$ and assume the weight to be fixed. This simplifies the derivation of $\hat{\boldsymbol{\eta}}$ as allowing $w(\hat{p}_{k^{\ast}}, \hat{p}_{k})$ to depend on the estimate of $\boldsymbol{\eta}$ requires advanced numerical techniques as the likelihood contains terms that require integration at each iteration.

\subsection{Atopic Dermatitis Studies}
This exploratory analysis studies the efficacy of an investigational drug $iT_x$ in adolescent patients aged 12-17 years with extrapolation of information from adult clinical trial. Data from an adolescent study (Study 2) and an adult study (Study 1) were included for these analyses. In Study 2, 150 patients were randomized 2:2:1 to $iT_x$ high dose, $iT_x$ low dose, or placebo. In Study 1, 300 patients were randomized 2:2:1 to $iT_x$ high dose, $iT_x$ low dose or placebo.

One endpoint of interest for both Study 1 and Study 2 was $\geq 75\%$ improvement from baseline in Eczema Area and Severity Index (EASI-75) at Week 12 and these were summarized in Table \ref{tab:propcrude}. The adolescent clinical trial information is the target population denoted by the subscript $k^\ast$ and the adult clinical trial is the reference population denoted by the subscript $k$. The response rates are denoted as $\hat{p}_{k^{\ast} j}$ and $\hat{p}_{k j}$, $j= 1, 2, 3$ for placebo, $iT_x$ low dose and $iT_x$ high dose, respectively.

Because there are two doses, the model (\ref{eq:workingmodel}) is extended to $\vartheta_{ik}=b_0 + b_1 I_{z_{ik}}(1) + b_2 I_{v_{ik}}(1) + \boldsymbol{x}_{ik}^{\top}\boldsymbol{\gamma}$, where $z_{ik}$ and $v_{ik}$ denote a binary covariate indicating the usage of $iT_x$ high dose (TRTHIGH: 1/0) and $iT_x$ low dose (TRTLOW 1/0) respectively. Other covariates in th emodel include baseline value of EASI (BASE: continuous), baseline severity measured by Investigator's Global Assessment categorized into `moderate' and `severe' (SEVERE: yes=1/no=0), where the coefficients for two covariates are denoted by $\gamma_1$ and $\gamma_2$ respectively.

For the bounded weight method, we derive $w_{placebo} = w(\hat{p}_{k^{\ast} 1}, \hat{p}_{k 1})$, $w_{low} = w(\hat{p}_{k^{\ast} 2}, \hat{p}_{k 2})$ and $w_{high} = w(\hat{p}_{k^{\ast} 3}, \hat{p}_{k 3})$ for placebo, $iT_x$ low dose and $iT_x$ high dose. The parameters for the weight function is specified the same as $w_1$ in the simulation study in Section \ref{sec:sim_study}. Three additional methods for no borrowing ($w_{placebo} = w_{low} = w_{high} = 0$), fixed weight borrowing ($w_{placebo} = w_{low} = w_{high} = 0.8$) and full borrowing ($w_{placebo} = w_{low} = w_{high} = 1$) are also included.

\begin{table}[!ht]
    \centering
     \begin{tabular}{l c c c c }
    \hline
        Study & ~ & Placebo & $iT_x$  & $iT_x$  \\ 
                ~ & ~ &  & Low QD & High QD \\ \hline
        Study 1 (adult) & N  & 61 & 125 & 114 \\
        ~ & n (\%) & 7 (11.5) & 46 (36.8) & 72 (63.2) \\ \hline
        Study 2 (adolescent) & N  & 29 & 55 & 66 \\
        ~ & n (\%) & 2 (6.9) & 25 (45.5) & 35 (53.0) \\ \hline
    \end{tabular}
    \caption{EASI-75 crude response rates in Study 1 and Study 2.   
}\label{tab:propcrude}
\end{table}

\begin{table}[!ht]
    \centering
    \begin{tabular}{l c c c c }
    \hline
        Method & Parameter & Estimation & 95\% CI  & P-value \\ \hline
        Bounded weights & $b_0$ &   -1.869 & (-2.582 ,  -1.155) & $<.$0001 \\ 
        $w_{placebo} = 0.8$, & BASE & -0.007 & (-0.014 ,  -0.001) & 0.0299 \\ 
        $w_{low} = 0.174$, & SEVERE & -0.475 & (-0.755 ,  -0.195) & 0.0009 \\ 
        $w_{high} = 0$ & TRTLOW &   1.974 & (1.250 ,   2.699) & $<.$0001 \\ 
        ~ & TRTHIGH &  2.470 & (1.722 ,   3.218 ) & $<.$0001 \\ \hline
        No borrowing & $b_0$ &  -2.282 & (-4.306 ,  -0.258) & 0.0271  \\ 
        ~ & BASE &  -0.005 & (-0.015 ,   0.006) & 0.3555 \\
        ~ & SEVERE &  -0.463 & (-0.833 ,  -0.093) & 0.0142 \\ 
        ~ & TRTLOW &  2.431 & (0.402 ,   4.460) & 0.0189\\ 
        ~ & TRTHIGH &  2.798 & (0.768 ,   4.828) & 0.0069 \\ \hline
        Fixed weights 0.8 & $b_0$ & -1.741 & (-2.426 ,  -1.057) & $<.$0001 \\ 
        ~ & BASE & -0.012 & (-0.017 ,  -0.007) & $<.$0001 \\ 
        ~ & SEVERE & -0.505 & (-0.659 ,  -0.351) & $<.$0001 \\ 
        ~ & TRTLOW &  1.817 & (1.130 ,   2.504) & $<.$0001 \\ 
        ~ & TRTHIGH &  2.704 & (2.007 ,   3.401) & $<.$0001  \\ \hline
        Full borrowing & $b_0$ &  -1.687 & (-2.341 ,  -1.032) & $<.$0001  \\
        ~ & BASE & -0.013 & (-0.018 ,  -0.008) & $<.$0001 \\ 
        ~ & SEVERE & -0.505 & (-0.652 ,  -0.358) & $<.$0001 \\ 
        ~ & TRTLOW &   1.773 & (1.116 ,   2.429) & $<.$0001 \\ 
        ~ & TRTHIGH &  2.706 & (2.040 ,   3.373) & $<.$0001 \\ \hline
    \end{tabular}
    \caption{Parameter estimations based on (composite) logistic regression model using various methods  
}\label{tab:abromle}
\end{table}

\begin{table}[!ht]
    \centering
    \begin{tabular}{l c c c c }
    \hline
        Method & Treatment Group & Response Rate & 95\% CI  & P-value  \\ \hline
        Bounded weights & Placebo & 0.093 & (0.034 ,  0.151) & ~\\ 
        $w_{placebo} = 0.8$, & $iT_x$ low dose & 0.417 & (0.382 ,  0.451) & ~ \\ 
        $w_{low} = 0.174$, & $iT_x$ high dose  & 0.537 &  0.478 ,  0.597) & ~ \\ 
        $w_{high} = 0$ & low dose vs Placebo & 0.324 & (0.255 ,  0.393) & $<.$0001 \\
        ~ & high dose vs Placebo & 0.445 & (0.361 ,  0.528) & $<.$0001 \\ \hline
        No borrowing & Placebo &  0.068 & (-0.059 ,  0.195) & ~ \\ 
        ~ & $iT_x$ low dose &  0.448 & (0.393 ,  0.502)  & ~ \\ 
        ~ & $iT_x$ high dose & 0.537 & (0.476 ,  0.598) & ~ \\ 
        ~ & low dose vs Placebo & 0.379 & (0.240 ,  0.519) & $<.$0001 \\ 
        ~ & high dose vs Placebo &0.469 & (0.328 ,  0.610)  & $<.$0001 \\ \hline
        Fixed weights .8 & Placebo & 0.091 & (0.036 ,  0.147) & ~ \\ 
        ~ & $iT_x$ low dose & 0.374 & (0.355 ,  0.393) & ~ \\ 
        ~ & $iT_x$ high dose & 0.585 & (0.547 ,  0.622) & ~ \\ 
        ~ & low dose vs Placebo & 0.282 & (0.223 ,  0.342) & $<.$0001 \\ 
        ~ & high dose vs Placebo & 0.493 & (0.426 ,  0.560) & $<.$0001 \\ \hline
        Full borrowing & Placebo &0.093 & (0.039 ,  0.147)  & ~ \\ 
        ~ & $iT_x$ low dose & 0.367 & (0.349 ,  0.385) & ~ \\ 
        ~ & $iT_x$ high dose & 0.589 & (0.552 ,  0.626) & ~ \\ 
        ~ & low dose vs Placebo & 0.275 & (0.217 ,  0.332)  & $<.$0001 \\ 
        ~ & high dose vs Placebo & 0.496 & (0.430 ,  0.561) & $<.$0001 \\ \hline
    \end{tabular}
\caption{Adolescent EASI-75 response rates based on (composite) logistic regression model using various methods     
}\label{tab:propmle}
\end{table}

The summary of parameter estimations, 95\% confidence intervals (CI) and p-values by Wald method are  summarized in Table \ref{tab:abromle}. For the bounded weight method, $w_{placebo} = 0.8$, $w_{low} = 0.174$ and $w_{high} = 0$ for placebo, $iT_x$ low dose and $iT_x$ high dose. $w_{placebo} = 0.8$ is the upper bound, as the crude response rates in $iT_x$ high dose for adult and adolescent clinical trials are similar. While $w_{low}$ and $w_{high}$ are close to 0, indicating not enough similarity under the current weight function thus minimal/none borrowing from the adult clinical trial for $iT_x$ doses. Of note, the bounded weight method has p-values that are between No borrowing and Fixed weight of 0.8 or Full borrowing, suggesting that it is a method that controls borrowing independent from the maximum weight derived from similarity in baselines. This is where the value of the methodology becomes clear even if the effect size observed is unequivocal. 

The coefficients for both treatment indicators for $iT_x$ low dose and $iT_x$ high dose are significant using different methods regardless of the extent of borrowing, representing the conditional treatment effect of $iT_x$ doses is higher than placebo group. For the covariates, the baseline value and the baseline disease severity both have marginally significant effects, except for the baseline disease severity when using no borrowing method.

Table \ref{tab:propmle} provides the estimated response rates and response differences for the adolescent population based on various methods of borrowing and using estimator proposed by \citep{freedman2008randomization, ge2011covariate}. For each participant in the adolescent study, the three model-based predictions of the probability of responses was calculated under three treatment groups (placebo, $iT_x$ low dose and $iT_x$ high dose) using each subject’s specific baseline covariates respectively. The response rates and response differences were calculated by averaging the predicted probability of responses. The standard errors were derived using a delta method. CI and p-values were based on Wald statistics and test. Of note, the length of the confidence interval of the bounded method is usually in between no borrowing and maximum borrowing again suggesting that it is a method that controls borrowing independent from the maximum weight derived from similarity in baselines.

Response rates for $iT_x$ low dose and $iT_x$ high dose are both significant higher than placebo using different methods regardless of the extent of borrowing. The point estimation for response rates using bounded weight method and no borrowing method are similar, the length of the confidence internal for placebo is lower when using bounded weight method (0.117) compared to the no borrowing method (0.255) reflecting the effect of borrowing in this treatment group ($w_{placebo} = 0.8$).

\subsection{Tipping Point Analysis and Effective Sample Size}
In the draft ICH E11A, one consideration is ``to understand {\it a priori} how much available information is being incorporated into the design and analysis to support the interpretation of the pediatric trial.'' The guidance goes on by saying that it is of relevance to know how much of the data that has been generated in the reference population is being used in the exercise, but also how much of the data generated in the reference population is relative to the amount of data that needs to be generated in the target population. Because the observed data can be highly variable due to many extrinsic factors in the trial, it is helpful if this can only be partly quantified {\it a priori} through some sort of what is the minimum borrowing as inferred from existing evidence on disease similarity and response intervention. This also justifies why an interval approach as discussed in Section \ref{specification}- \ref{bounded_single} is appropriate. Otherwise, everything is {\it post facto} and there will be cases where, due to variability, the reference data is inadequate as additional information to the newly generated data in the target population to support conclusions of efficacy and safety. 

A ``tipping-point'' analysis is then needed to measure how much the reference data is informing the study conclusion. In particular, it is needed to understand how conclusions change with the amount of borrowing used relative to the data generated, i.e., how the reference data is driving the estimation of the effects in the target population. The larger the nominal information used is, the more unlikely the conclusion of efficacy in the target population is true or that the currently generated data is inadequate to provide unequivocal conclusions about the efficacy and safety of the drug in the target population. 

To perform this tipping point analysis, we calculated what is the effect of using the reference population given varying weights on the variance of the estimates assuming a similar effect in the reference population. We also performed an analysis of when responses between reference and target differ and how the bounded weight can help in this situation. The idea is to check at what point will the data generated in the target population be rendered inadequate and how the bounded interval can put a ``stop gap'' on this uncontrolled variability. In other words, the use of an interval for the weight gives some lever that the extent of development will be sufficient.
Revisiting the pediatric atopic dermatitis study, the results of testing the efficacy of the treatment does not change with the change in weight.
Thus, there exists no tipping point for this specific example.

We derive an effective sample size measuring the number of reference observations borrowed by adopting the guidance of Bayesian statistics presented by FDA as below.
\begin{equation}
    ESS = n_{k^\ast}\left(\frac{Var(\hat{\theta}_j|D_t)}{Var(\hat{\theta}_j|D_r, D_t)}-1\right),
\end{equation}
where $\hat{\theta}_j$ is the $j^{th}$ parameter of interest.
For the pediatric atopic dermatitis study, when the weight is 0.8, the effective sample sizes for parameters $\hat{b}_1$, $\hat{b}_2$, $\hat{\gamma}_1$, and $\hat{\gamma}_2$ are 36391, 719, 942, and 1025 respectively, which are greater than the number of reference observations.
The effective sample sizes are zero when we specify the weight as zero.

\section{Discussion}
Extrapolation provides the scientific justification for the use of aggregative analytic methods, e.g., information borrowing/bridging through Bayesian and meta-analytic methods, for obtaining efficacy conclusions in children. This in itself requires thoughtful discussion and implementation which are elucidated in the following:  

{\bf Scientific and Methodological Alignment}. One key concept in the use of these innovative analytical strategies is whether the methodological construct aligns with the scientific framework. The regulatory decision to require studies in children is an acknowledgment of the potential benefit for children. However, the implementation between some extrapolation and no extrapolation is highly dependent on a single outcome of the pediatric trial. In reality, extrapolation requires the assessment of the similarity of disease and an assessment of the similarity of treatment response beforehand. If extrapolation is based purely on outcomes, then it will be subject to chance findings and other post-randomization events that impact the study. Penalizing for heterogeneity of outcomes that are a product of several post-randomization factors as opposed to the scientific determination of the similarity of the disease seems counter-intuitive. Furthermore, there may be more variability in one endpoint than another and hence the amount of information to be borrowed will vary from one endpoint to another. These considerations suggest having a nuanced statistical methodology to ensure they are aligned with the concept scientifically. Furthermore, having a methodology that aligns with the science makes for an easier justification of the proposal in regulatory submission and interpretation of the results of the trial. 

{\bf Controlled borrowing within Composite Likelihood}. Extrapolation goes beyond variability in response, i.e. the differences in response to intervention do not invalidate extrapolation. Hence, the leveraging of information can be pre-specified through a similarity coefficient $\alpha\in [a,b]\subset[0,1]$, which is dependent on an evidence-based ``line of reasoning'' established before trial conduct and agreed upon with regulatory agencies. This composite score can explore several factors which can then be aggregated in terms of single domain scoring corresponding to some general arguments on the prior belief that the drug has no relevant effect when a standard drug development program has already passed Phase 2 and has arrived to plan Phase 3. 

The observation that differences in response to intervention do not invalidate extrapolation also implies that even if there are some minor differences in the treatment effect, the use of maximum borrowing is still warranted, i.e., $\alpha = b$. Whereas, even if the treatment effect is quite different but as long as benefit is still achieved then a minimum borrowing instead of no borrowing can be allowed, i.e, $\alpha=a$. This is, indeed, the essence of extrapolation as described in ICH E11A. We have provided three weight functions that achieve these desired characteristics and have also offered how to determine the weight in the case of multiple treatment groups.  

The use of composite likelihood also can use linear models that incorporate covariate adjustment. Maximum likelihood approaches are then used for estimation of parameters and asymptotic theory is used to derive distributions of estimates for use in inference. This type of borrowing also easily incorporates missing data methods. 

{\bf Type I error and Tolerable Uncertainty}. The nature of extrapolation is an acknowledgment of a different view of tolerable uncertainty, that is, what is the acceptable risk of having a wrong decision over the approval of an investigational drug? Efficacy (from an FDA regulatory perspective) requires either two clinical trials, one clinical trial with overwhelming evidence, or one clinical trial with additional supporting evidence (21 CFR 514.4). Most pediatric drug developments do not require this same level of certainty. The decision to expose a child to the risks of a new drug in a research protocol is already framed by the risks and prospect of direct benefit to that individual child. Hence, when a drug is known to be effective in adults, there should be a willingness to incorporate this information into the pediatric program planning, regardless of the statistical approach, which will necessarily increase type 1 error. The more similar the diseases are, the higher the confidence that the type I error, which is a concept defined in the null hypothesis, needs to be modified. An argument can then be made that type I error may not be the appropriate tool to measure false positive decisions in pediatric drug development.

\newpage
\bibliographystyle{unsrt}  
\bibliography{reference}

\end{document}